\documentclass[sigconf,nonacm]{acmart}
\usepackage{multirow}
\usepackage{subfigure}
 \usepackage{url}
 \usepackage{multirow}

\usepackage{algorithmic}
\usepackage{graphicx}
\usepackage{xcolor}
\usepackage{soul}

\AtBeginDocument{%
  }

\setcopyright{none}

\newcommand\xor{\oplus}

\graphicspath{{Figures/}{/home/bman/dev/logos/}}%

\begin{document}
\title{A Statically and Dynamically Scalable Soft GPGPU}

\author{Martin Langhammer}
\affiliation{%
  \institution{Intel Corporation \& Imperial College London}
  \country{London, UK}
}
\email{martin.langhammer@intel.com}

\author{George A. Constantinides}
\affiliation{%
  \institution{Imperial College London}
  \country{London, UK}
}
\email{g.constantinides@imperial.ac.uk}

\begin{abstract}
Current soft processor architectures for FPGAs do not utilize the potential of the massive parallelism available. FPGAs now support many thousands of embedded floating point operators, and have similar computational densities to GPGPUs. Several soft GPGPU or SIMT processors have been published, but the reported large areas and modest Fmax makes their widespread use unlikely for commercial designs. In this paper we take an alternative approach, building the soft GPU microarchitecture around the FPGA resource mix available. 
We demonstrate a statically scalable soft GPGPU processor (where both parameters and feature set can be determined at configuration time) that always closes timing at the peak speed of the slowest embedded component in the FPGA (DSP or hard memory), with a completely unconstrained compile into a current Intel Agilex FPGA. We also show dynamic scalability, where a subset of the thread space can be specified on an instruction-by-instruction basis.

For one example core type, we show a logic range -- depending on the configuration -- of 4k to 10k ALMs, along with 24 to 32 DSP Blocks, and 50 to 250 M20K memories. All of these instances close timing at 771 MHz, a performance level limited only by the DSP Blocks. We describe our methodology for reliably achieving this clock rate by matching the processor pipeline structure to the physical structure of the FPGA fabric. We also benchmark several algorithms across a range of data sizes, and compare to a commercial soft RISC processor. 
\end{abstract}

\maketitle

%%%%%%%%%%%%%%%%%%%%%%%%%%%%%%%%%%%%%%%%%%%%%%%%%%%%%%%%%%%%%%%%%%%%%%%%%%%
\section{Introduction}
%%%%%%%%%%%%%%%%%%%%%%%%%%%%%%%%%%%%%%%%%%%%%%%%%%%%%%%%%%%%%%%%%%%%%%%%%%%

FPGAs are capable platforms, with multiple thousands of embedded memories as well as DSP Blocks, many of which now support IEEE 754 floating point numerics. In addition, there is a significant amount of high performance IP available for FPGAs, {\em{e.g.}}~FFTs~\cite{IntelFFT,Xilinx_FFT} and error correction such as Reed-Solomon codecs~\cite{IntelRS}. High performance systems can readily be assembled using a combination of original design and these IP Blocks. The value of the FPGA is integration: although each individual IP or function is lower performance than ASIC, this is offset by the flexibility. However, modifying IP - even your own - requires significant effort. FPGA hardware compile times (synthesis, place and route) can take hours, and timing closure can be a significant unknown. Implementing (and modifying) a complex subset of a system by a pure software approach, where the result of the compile or assembly is essentially instantly available, and loaded onto an already placed and routed processor, may be very attractive.   

Soft RISC cores (Nios~\cite{NiosV} and MicroBlaze~\cite{Microblaze}) for FPGA have been used for over two decades, and allow the inclusion of complex control flow, or the offload of ancillary functions. Although these RISC processors are very flexible, they also have a rather low performance. Parallel processor architectures may offer better performance, and SIMT (GPGPU) processors may be able to efficiently use the large number of memory and DSP Blocks distributed across the FPGA device. There have been a number of soft SIMT FPGA architectures published ~\cite{FlexGrip,MIAOW,FGPU,SCRATCH,DOGPU,Guppy,Kingyens}, but these are often very large (50K-300K LUTs), and typically have a low clock frequency (30MHz-100MHz). Other types of parallel processors are also known for FPGA ~\cite{Vegas,Venice,VectorBlox} (and commercialized ~\cite{MicrosemiVectorBlox}), but the Fmax is relatively low at $\sim$150MHz. 

A different approach has been taken by Xilinx (now AMD) in the Versal devices, with arrays of AI Engines, a hardened VLIW processor. This motivates us to consider whether we can combine the flexibility of a soft processor (where any number can be instantiated into the soft fabric), but with the performance of an ASIC implementation (in this case, running at the speed of the embedded hardened features).

Our design, which we call the eGPU (for {\em{embedded}}GPU), is both statically and dynamically scalable, features which make it particularly useful and performant for FPGA applications. Static scalability is the ability to parameterize the thread space, shared memory space, integer ALU functions, as well as major processor features (such as predicates). Dynamic scalability allows us to operate on a defined subset of the thread space, and change this on an instruction by instruction basis, without any dead time. We will see that this can greatly reduce the number of cycles required in some portions of the program, such as during a vector reduction (which is a common kernel of GPGPU applications).   

We make the following contributions:
\begin{itemize}
    \item Describe a novel parameterized SIMT processor for FPGA, with a wide range of user defined instructions, as well as architectural trade-offs (such as predicates).
    \item Demonstrate that a soft processor can consistently close timing at a level limited only by the embedded features such as DSP and memory, and do so with a completely unconstrained compile.
    \item Compare the absolute and normalized (by resource cost) results of a soft GPGPU with a soft RISC processor, and show that the SIMT architecture is better in the general case, and significantly better when using dedicated hardware extensions.
\end{itemize}

\section{Background}

Our goal for this project was to architect and implement a compact, high performance SIMT processor, that can be used for commercial FPGA system designs. We can use current and prior FPGA processors both to understand the limitations of previous projects, and to validate some of our design choices. The axes of comparison to other work include memory systems, complexity (such as workload balancing), and trade-offs between hard and soft implementation.

Many of the previously published GPGPUs~\cite{FlexGrip,FGPU,DOGPU,SCRATCH} are SIMT processors which were compiled to an FPGA, whereas eGPU was designed {{\em for}} FPGA. The eGPU has an power-performance-area (PPA) metric which is one or two orders of magnitude (OOM) smaller than some of the earlier soft GPGPUs. Comparisons between high-performance processor designs are complex and multi-dimensional. For example, some existing soft GPUs have more complex memory systems, including caches and dynamic workload balancing. This does come with a cost, with a typical order of magnitude resource difference, as can be seen in Table~\ref{tab:table_softgpus}, where we compare configurations of the other soft GPGPUs that are closest in computational structure to eGPU (PEs are roughly the same as SPs). Despite the much deeper pipelines ({\em{e.g.}}~FlexGrip~\cite{FlexGrip} has a 21 deep pipeline, FGPU has a 18 deep pipeline~\cite{FGPU}), they also run at a considerably slower clock frequency. Although they are implemented in older FPGA technology (FlexGrip is in Virtex-6 at 100MHz), this does not fully explain the performance level, as there are soft processors that run at 450MHz in those devices~\cite{cheahFPT}~\cite{Cheah}. In the benchmarking section we will also see that the benchmarks also run slower than expected on the earlier GPGPUs based on the difference in clock frequency.

\begin{table}
\footnotesize
  \begin{center}
    \caption{Resource Comparison}
    \label{tab:table_softgpus}
    \begin{tabular}{|c|c|c|c|c|c|c|} 
    \hline
       \textbf{Architecture} & \textbf{Config.} & \textbf{LUTs}& \textbf{DSP} & \textbf{FMax} & \textbf{PPA} & \textbf{Device} \\
      \hline \hline
      \textbf{FGPU~\cite{FGPU}} & {2CUx8PE} & {57K} & 48 & 250 & {36} & {Zynq-7000}\\
        \hline
      \textbf{DO-GPU~\cite{DOGPU}} & {4CUx8PE} & {360K} & 1344 & 208 & {133} & {Stratix 10} \\
        \hline
      \textbf{FlexGrip~\cite{FlexGrip}} & {1SMx16PE} & {114K} & 300 & 100 & {175} & {Virtex-6} \\
        \hline
      \textbf{eGPU} & {1SMx16SP} & {5K} & 24 & 771 & {1} & {Agilex} \\
      \hline
    \end{tabular}
  \end{center}
\end{table}

Instead, we validate eGPU against existing soft RISC processors~\cite{NiosV}, which are extensively used in real applications. We will normalize the benchmark results based on cost {{\em i.e.}}~FPGA resources consumed. The eGPU, being a parallel processor (with essentially 16 smaller multi-threaded processors) will naturally be larger; to be effective and usable, it must have a clear advantage in both absolute performance and normalized efficiency over the RISC processors.  

\begin{figure*}
    \centering
    \includegraphics[scale=0.83]{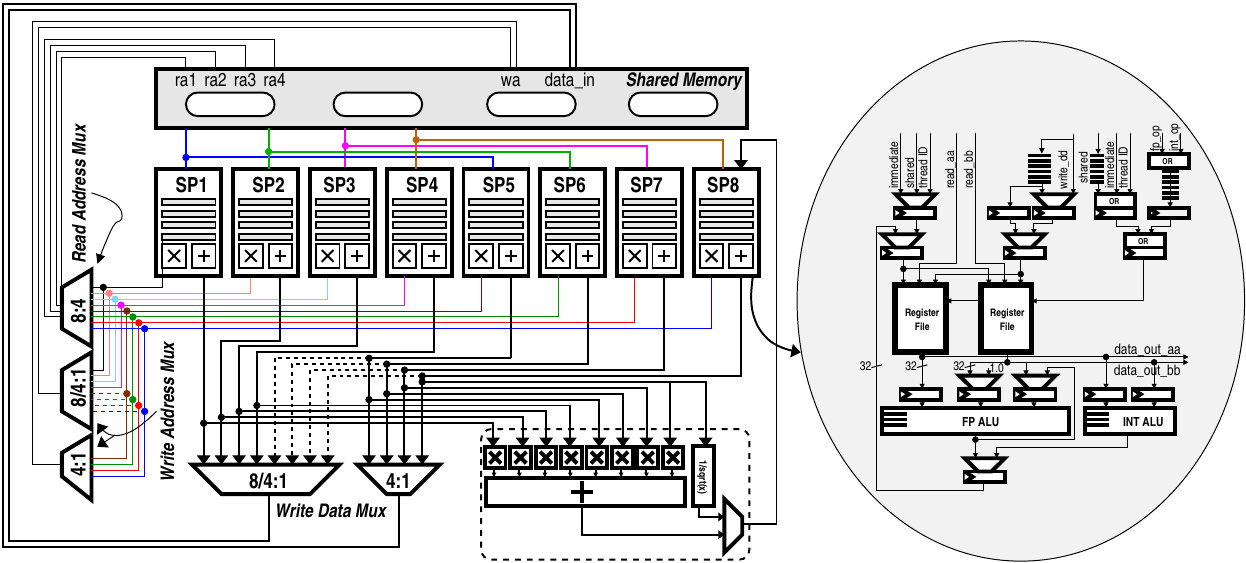}
    \caption{eGPU SM Top Level Architecture}
    \label{fig:SM_SP_Top}
\end{figure*}

eGPU uses a single local data memory, which is configurable in size, and does not support a cache. Larger datasets need to be externally managed. Like the eGPU, the Xilinx AI Engines~\cite{XilinxAIE}, which are organized as hard VLIW hard processor arrays, have only a single local data memory per CPU, the loading and unloading of which has to be managed externally. Algorithms with larger amounts of data (such as 4K FFTs) need to be split across multiple AI Engines~\cite{XilinxAIEFFT}. The eGPU has a greater memory flexibility, as we are able to configure a larger shared memory instance (we show examples with up to 128KB in this paper). The AI Engines give us an example of a commercial FPGA parallel processor, where using multiple simpler processors have been found to have an advantage over using complex memory systems. 

%%%%%%%%%%%%%%%%%%%%%%%%%%%%%%%%%%%%%%%%%%%%%%%%%%%%%%%%%%%%%%%%%%%%%%%%%%%
\section{Architecture Description}
%%%%%%%%%%%%%%%%%%%%%%%%%%%%%%%%%%%%%%%%%%%%%%%%%%%%%%%%%%%%%%%%%%%%%%%%%%%

The architecture of the eGPU is based on an earlier proof-of-concept design~\cite{eGPU_FPL}.
Our new design adds significant scalability - thread and register space, shared memory size, instruction set support, as well as optional predicates for thread divergence. Figure~\ref{fig:SM_SP_Top} shows the top level architecture of the eGPU. The streaming multi-processor (SM) contains 16 parallel scalar processors (SP), although only 8 are shown in the figure for clarity. An optional dot-product core and special function unit (SFU) reciprocal square root can be attached. We target the Intel Agilex~\cite{ChromczakAgilex} family of FPGAs in this work. The eGPU has a very short pipeline (8 stages) compared to other GPUs; therefore, hazards are hidden for most programs. Consequently, we do not provide hardware support for tracking hazards in the current version, which in turn gives us an efficient and fast processor. 

Two types of embedded memories are now supported, simple dual port (DP) and the emulated quad port (QP) blocks~\cite{Agilex_Memory}. One of the largest performance limitations of the earlier eGPU architecture was memory bandwidth. The QP memory will double the write bandwidth, while at the same time reducing the number of embedded memory blocks required (the 20K-bit M20K blocks) by half. The trade-off is that in QP mode, the memory speed is reduced from 1 GHz to 600 MHz, which then becomes the critical path in the processor. Resource, Fmax, and benchmark results are all described later in this paper.

\subsection {Dynamic Scalability}

 Most GPGPUs support thread divergence by predicates (thread-specific conditionals) but these have a potential significant performance impact, as all threads are run, whether or not they are written back. In addition to predicates, the eGPU sequencer supports an instruction by instruction specification of a subset of the thread space, where only the indicated threads are run.  If the program can be constructed such that the data of interest can be written to the threads that can be isolated by the dynamic thread allocation, then a large number of processing cycles can be skipped. This is particularly noticeable in programs with many multi-cycle instructions, such as reads and writes to shared memory. This will have a direct impact on the benchmark performance (number of cycles).

We define a wavefront as the maximum number of operations that can be run per clock cycle; with 16 SPs we have a wavefront width of 16. The thread block depth (alternately, the wavefront depth) is the number of wavefronts per instruction, which is the initialized thread size / 16. We feel these terms allow us to describe our dynamic thread scalability more concisely.

The eGPU can be configured, on a cycle by cycle basis, to act as a standard SIMT processor, a multi-threaded CPU, or a single threaded MCU. While the number of clock cycles to execute all the threads for an operation instruction ({{\em e.g.}~}FP or INT) is dependent on the depth of the thread block, loads and stores are multi-cycle (because of the limited number of ports to shared memory).  The impact of dynamically adjusting the width of certain instructions ({{\em e.g.}~}reduction, where the writeback data can be orders of magnitude less than the read data) can be seen in the benchmark section later in this paper. 

The upper 4-bit field in the instruction word (IW) allows the wavefront width and depth to be coded for that instruction. Perhaps the most common case will be using only the first SP, or even the first thread in the first SP; many GPU applications will have vector reduction kernels, where a reduction result(s) may end up in the leftmost SP. If we can operate on this SP exclusively for a certain subset of time during the execution of the program, we can save significant processing time, and power. The coding of the personality is described in the Instruction Set section.

Hence, we can change the scale of the SIMT span by reducing the wavefront width and/or depth. The eGPU can act as a multi-threaded CPU if we set the wavefront width to one, and if we also set the thread depth to one, each instruction will only act on thread 0 of the first SP - this SP can then be used like a MCU. We will use these modes to good effect in our benchmarks later in this paper.  

\subsection {Predicates}

The eGPU optionally - by user configuration - supports predicates, which enable thread divergence. Conditionals can be applied to each thread individually using a condition instruction (see Table~\ref{tab:instruction_set}).
 
As with many aspects of eGPU, the number and type of conditions can be selected at compile time. Although these will have only a minimal impact on area, the additional wireload may impact performance because of the large number of individual predicate stacks. There is one predicate stack per initialized thread, so there may be thousands of stacks per eGPU instance.  

Some algorithms, such as the bitonic sort benchmark in this paper, require predicates. On the other hand, many of the signal processing applications that we expect that the eGPU will be used for (such as FFTs and matrix decomposition) do not use data dependent decisions. These do not need predicates, and can be programmed using only loop constructs, which are supported in the eGPU sequencer. For this reason, the presence and complexity of predication is a parameter of our design, especially considering the large potential cost of the feature.

\begin{figure}
    \centering
    \includegraphics[scale=0.40]{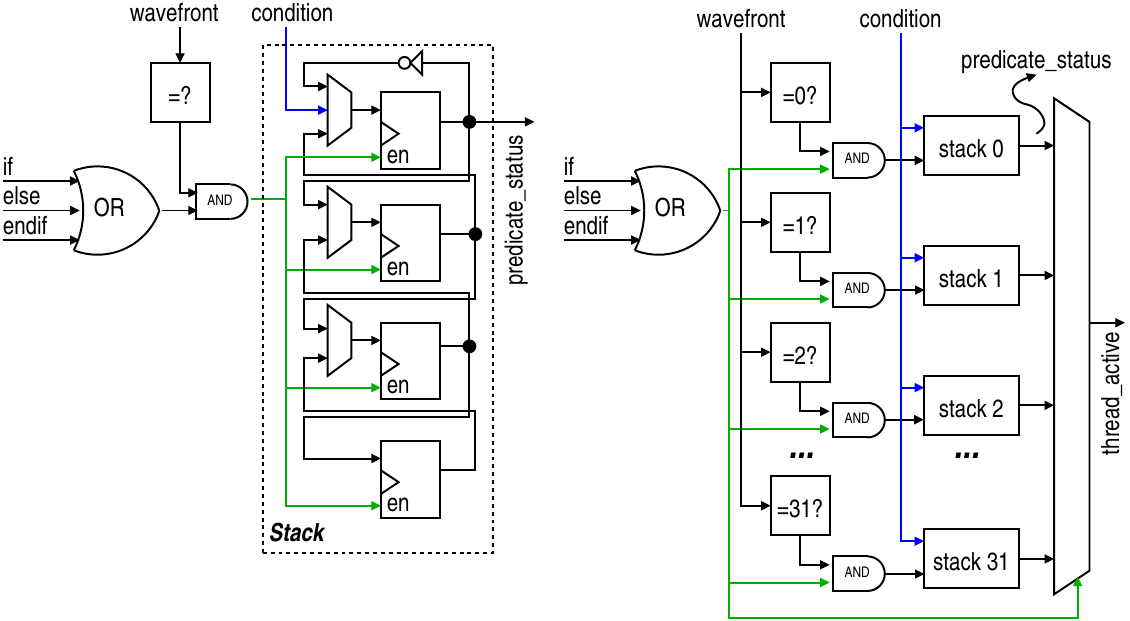}
    \caption{One Predicate Block}
    \label{fig:predicate_stack}
\end{figure}

Figure ~\ref{fig:predicate_stack} shows the structure of a single predicate block. Each SP has a separate block, which are comprised of multiple predicate stacks. Each thread has a unique predicate stack. Multiple nested levels of conditional operations (IF/ELSE/END IF) are supported per stack, with the maximum supported depth of nesting being parameterized. 

The incremental cost of adding one level of nesting is trivial, as the control logic of each predicate stack is the dominant user of logic here. The wavefront value (for example, in our base eGPU configuration of 512 threads with 16 SPs, there will be 32 wavefronts - {{\em i.e.}~}32 threads per SP) enables the correct predicate stack for the current thread. If the condition instruction (IF) condition is true for that thread, a '1' will be set at the top of the predicate stack, and the rest of the stack pushed down. An ELSE instruction will invert the top of the stack, and an END IF will pop the stack and return to the previous nesting level. 

 The eGPU is configured at compile time for a maximum number of threads; if the run time configuration of threads is less than this, there is no issue as only the selected threads will trigger the operation of the predicate block. 

The conditional value will only be applied to the current predicate block, and all others ignored in that clock cycle. The current thread activation 
{\texttt{thread\_active}} signal will be muxed from all the predicate blocks, selected by the current wavefront. The {\texttt{thread\_active}} signal is used to pass or zero the {\texttt{write\_enable}} signals to either the register files or shared memory, whichever is the destination for that instruction.

%%%%%%%%%%%%%%%%%%%%%%%%%%%%%%%%%%%%%%%%%%%%%%%%%%%%%%%%%%%%%%%%%%%%%%%%%%%
\section{Instruction Set}
%%%%%%%%%%%%%%%%%%%%%%%%%%%%%%%%%%%%%%%%%%%%%%%%%%%%%%%%%%%%%%%%%%%%%%%%%%%

Table ~\ref{tab:instruction_set} shows most of the instruction set for the eGPU. There are a total of 61 instructions, including 18 conditional cases (we omit the FP conditional instructions here for brevity). Usually, only a subset of instructions are included (by the user defined configuration of the eGPU). The 18 conditional cases depend on predicates being included in the parameters - as predicates typically increase soft logic cost by 50\% they are only used when the expected class of applications need them. Many of the intended applications, such as FFT, matrix multiplication and decomposition, do not, and the required loops can be handled with the dedicated loop instructions. Some instructions can support multiple {\em{TYPES}}, such as signed {\em{(INT32)}} and unsigned {\em{(UINT32)}} formats for integer instructions.

\begin{table}
\footnotesize
  \begin{center}
    \caption{Instruction Set}
    \label{tab:instruction_set}
    \begin{tabular}{|c|l|l|}
    \hline
    \textbf{Group} & \centering \textbf{Instruction} & \textbf{Operation} \\
    \hline \hline
    \multirow{4}{*}{Integer Arithmetic} & \texttt{ADD.TYPE Rd,Ra,Rb} & {Rd = Ra + Rb} \\
    \ & \texttt{SUB.TYPE Rd,Ra,Rb} & {Rd = Ra - Rb}\\
    \ & \texttt{NEG.TYPE Rd,Ra} & {Rd = -Ra}  \\
    \ & \texttt{ABS.TYPE Rd,Ra} & {Rd = absolute(Ra)}  \\
    \hline
    \ \multirow{4}{*}{Integer Multiply} & \texttt{MUL16LO.TYPE Rd,Ra,Rb} & 
    {Rd = Ra * Rb} \\
    \ & \texttt{MUL16HI.TYPE Rd,Ra,Rb} & {Rd = (Ra * Rb)>>16}  \\
    \ & \texttt{MUL24.LO.TYPE Rd,Ra,Rb} & {Rd = Ra * Rb}  \\
    \ & \texttt{MUL24.HI.TYPE Rd,Ra,Rb} & {Rd = (Ra * Rb)>>24}  \\
    \hline
    \multirow{6}{*}{Integer Logic} & \texttt{AND Rd,Ra,Rb} & {Rd = Ra \& Rb} \\
    \ & \texttt{OR Rd,Ra,Rb}  &  {Rd = Ra $\|$ Rb} \\
    \ & \texttt{XOR Rd,Ra,Rb}  & {Rd = Ra $\xor$ Rb} \\
    \ & \texttt{NOT Rd,Ra} & {Rd = !Ra} \\
    \ & \texttt{cNOT Rd,Ra} & {Rd = (Ra == 0)?1:0} \\
    \ & \texttt{BVS Rd,Ra} &  {Rd = bit\_reverse(Ra)} \\
    \hline
    \multirow{2}{*}{Integer Shift} & \texttt{SHL.TYPE Rd,Ra,Rb} &  {Rd = Ra $\ll$ Rb} \\
    \ & \texttt{SHR.TYPE Rd,Ra,Rb} &  {Rd = Ra $\gg$ Rb} \\
    \hline
    \multirow{3}{*}{Integer Other} & \texttt{POP Rd,Ra} &  {Rd = unary(Ra)} \\
    \ & \texttt{MAX.TYPE Rd,Ra,Rb} & {Rd = (Ra>Rb)?Ra:Rb} \\
    \ & \texttt{MIN.TYPE Rd,Ra,Rb} & {Rd = (Ra<Rb)?Ra:Rb} \\
    \hline
    \multirow{7}{*}{FP ALU} & \texttt{ADD.FP32 Rd,Ra,Rb} & {Rd = Ra + Rb} \\
    \ & \texttt{SUB.FP32 Rd,Ra,Rb} & {Rd = Ra - Rb}  \\
    \ & \texttt{NEG.FP32 Rd,Ra} & {Rd = -Ra}  \\
    \ & \texttt{ABS.FP32 Rd,Ra} & {Rd = absolute(Ra)}  \\
    \ & \texttt{MUL.FP32 Rd,Ra,Rb} & {Rd = Ra*Rb}  \\
    \ & \texttt{MAX.FP32 Rd,Ra,Rb} & {Rd = (Ra>Rb)?Ra:Rb}\\
    \ & \texttt{MIN.FP32 Rd,Ra,Rb} & {Rd = (Ra<Rb)?Ra:Rb}\\
    \hline
    \multirow{6}{*}{Int Compare} & \texttt{eq} & {$Ra == Rb$} \\
    \ & \texttt{ne} & {$Ra \ne Rb$} \\
    \ & \texttt{lt (INT), lo (UINT)} & {$Ra < Rb$} \\
    \ & \texttt{le (INT), ls (UINT)} & {$Ra \leq Rb$} \\
    \ & \texttt{gt (INT), hi (UINT)} & {$Ra > Rb$} \\
    \ & \texttt{ge (INT), hs (UINT)} & {$Ra \geq Rb$} \\
    \hline
    \multirow{2}{*}{Memory} & \texttt{LOD Rd (Ra)+offset} & {Read from Shared} \\
    \ & \texttt{STO Rd (Ra)+offset} & {Write to Shared}   \\
    \hline
    \ {Immediate} & \texttt{LOD Rd \#Imm} & {Rd = Imm} \\
    \hline
    \multirow{2}{*}{Thread} & \texttt{TDx Rd } & {Rd = Thread IDx} \\
    \ & \texttt{TDy Rd} & {Rd = Thread IDy} \\
    \hline
    \multirow{3}{*}{Extension} & \texttt{DOT Rd,Ra,Rb} & {Dot Product $\langle Ra,Rb \rangle$} \\
    \ & \texttt{SUM Rd,Ra,Rb} & {Reduction $\langle Ra,Rb \rangle$} \\
    \ & \texttt{INVSQR Rd,Ra} & {$Rd = 1/\sqrt {Ra}$} \\
    \hline
    \multirow{6}{*}{Control} & \texttt{JMP address} & {Jump to Address} \\
    \ & \texttt{JSR address} & {Subroutine Address} \\
    \ & \texttt{RTS} & {Return from Subroutine} \\
    \ & \texttt{LOOP address} & {Jump and Dec Loop Ctr} \\
    \ & \texttt{INIT loops} & {Set Loop Ctr} \\
    \ & \texttt{STOP} & {Stop and Set Flag} \\
    \hline
    \multirow{3}{*}{Conditional} & \texttt{IF.cc} & {if cc true} \\
    \ & \texttt{ELSE} & {if cc false} \\
    \ & \texttt{ENDIF} & {clear cc} \\
    \hline
    \end{tabular}
  \end{center}
\end{table}

The integer ALU uses a large proportion of the soft logic ($\approx$100 ALMs to $\approx$400 ALMs), so selecting only the required precision (16-bit or 32-bit) and feature subset can reduce the cost of the eGPU substantially. Extension instructions are also optional. We will use the dot product instruction for some of the benchmarks in this paper; if used, it can make significant difference to the performance of some functions. We can also add elementary functions (currently we support only reciprocal square root), which are required for algorithms such as matrix decomposition. In contrast, the FP instructions are almost completely contained inside the DSP Block, with only the FP {\em{Max()}} and {\em{Min()}} instructions having a potential impact on area or performance. 

\begin{figure}
    \centering
    \includegraphics[scale=0.55]{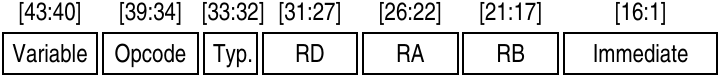}
    \caption{Instruction Word}
    \label{fig:iword}
\end{figure}

Figure ~\ref{fig:iword} shows an instruction word, here shown in a 43-bit form. As the number of registers per thread changes, the three register field widths also change; the displayed word is for a 32 registers per thread configuration, which requires 5 bits to encode the register number. The 2-bit representation field encodes whether the number is unsigned integer, signed integer, or FP32. The four most significant bits encode the processing type, which allow the wavefront depth and the width of the wavefront to be changed on an instruction by instruction basis. 

Writing these results into shared memory using subset write can be 16x faster than using the generic write. An instruction, whether used for a full or a partial thread space, is almost identical, with only the four instruction type bits used to control the subset of the thread space. Table ~\ref{tab:subset_thread_iw} shows how the upper 4 bits of the IW control the width and depth of the thread space.

\begin{table}
\small
  \begin{center}
    \caption{Thread Space Control}
    \label{tab:subset_thread_iw}
    \begin{tabular}{|c|c|c|}
    \hline
    \textbf{Coding} & \textbf{Width [4:3]} & \textbf{Depth [2:1]}  \\
    \hline\hline
    \ {"00"} & {All (16 SPs)} & {Wavefront 0 only} \\
    \hline
        \ {"01"} & {1/4 width (first 4 SPs)} & {all wavefronts} \\
    \hline
        \ {"10"} & {SP0 only} & {first 1/2 wavefronts} \\
    \hline
        \ {"11"} & {Undefined} & {first 1/4 wavefronts} \\
    \hline
    \end{tabular}
  \end{center}
\end{table}

%%%%%%%%%%%%%%%%%%%%%%%%%%%%%%%%%%%%%%%%%%%%%%%%%%%%%%%%%%%%%%%%%%%%%%%%%%%
\section{Results}
%%%%%%%%%%%%%%%%%%%%%%%%%%%%%%%%%%%%%%%%%%%%%%%%%%%%%%%%%%%%%%%%%%%%%%%%%%%
We compiled a number of different eGPU instances, using both DP and QP memory versions. We used Quartus Pro V22.4 and targeted an Intel Agilex AGIB027R29A1E1V device~\cite{AgilexAGF027}. All of our results are reported for a single compilation attempt (we did not use seed sweeps).
 
The DP memory results are tabulated in Table~\ref{tab:dp_fitting}. We define three categories - small, medium, and large - to show the effects of different thread space, shared memory, and ALU features, as well as the impact of supporting predicates. The base eGPU architecture is the same for all instances: one SM with 16 SPs, a two read port register memory, and a four read and one write port shared memory. We configured all of these cases to use 512 threads, but with varying numbers of registers per thread. QP memory results are shown in Table~\ref{tab:qp_fitting}, the main architectural change being the two write port shared memory.

The `small' category uses a 16-bit ALU, which will likely only be used for address generation. The minimum specification supports only a single bit shift, as well as a 16-bit adder/subtractor, and arithmetic logic (AND/OR/XOR) operations. The memory requirements for the SPs is reduced by providing 16 registers per thread. The `large' category implements 64 registers per thread, and larger shared memory sizes with up to 128KB. The integer ALU supports the full set of integer instructions defined in the previous section. We also include a `medium' category for further examples. Many other combinations of parameters and features sets are possible as well.

\subsection {Impact of Register and Shared Memory Size}

Both the thread registers and the shared memories are implemented using M20K memories, which can be configured into either DP (one read port and one write port active simultaneously) or QP memories (two read ports and two write ports active simultaneously). The natural datapath size of the eGPU is 32-bits, defined by the native IEEE 754 single precision (FP32) floating point DSPs which will be doing the majority of the calculations. In DP mode, a M20K can be configured as a 512x32-bit memory. Port restrictions mean that in QP mode the M20K is a 2048x8-bit block, which requires a larger minimum thread register space to take advantage of the extra ports. 

In DP mode thread registers are implemented in two dual port memories, providing two read ports and one write port per clock cycle. In our most common eGPU configuration (with 16 registers per thread), a 512 thread machine will require two M20Ks per SP, or 32 M20Ks for thread registers in total, which is also the minimum size. Both the number of registers per thread and the number of total threads are parameterized, but the number of M20Ks will increase accordingly. In QP mode, the 8-bit data port width means that there is no point in using less than 2$k$ thread registers per SP, although we will use half the number of M20K blocks compared to the DP version when we configure at least these number of registers.

The shared memory is implemented as a four read port, one write port per memory in DP mode. The smallest possible shared memory is 512 words (2KB), which would require four M20Ks. This is very small, and unlikely to be useful, as the shared memory size would only be as large as the register space in a single SP. A more realistic shared memory size would be 2$k$ words (8KB), which would require 16 M20Ks; the total memory usage for a small eGPU instance, including registers, would therefore be 48 M20Ks. The shared memory is set by parameter, and significantly larger sizes are possible without frequency impact. For example, a 64KB shared memory needs 128 M20Ks, and a 128KB shared memory 256 M20Ks, which is a small fraction of the memories on the device. In QP mode, the number of M20Ks is halved, and the number of write ports doubled to two. 

\subsection {Integer ALU Architecture and Resources}

Unlike the floating point arithmetic, which can be mapped directly to a DSP Block, the simpler integer operations need to be implemented in soft logic. We will see that up to half of the soft logic and registers in an eGPU is required for the integer ALU. Table~\ref{tab:alu_fitting} shows the resources, split by operation type, for a wide range of integer ALUs.

The smallest reasonable integer ALU is a 16 bit version with single bit shifts, which consumes 90 ALMs and 136 registers, most of which are used for the 5 stage pipeline. Here we have a signed adder/subtractor, as well as logic functions (in this case, only AND, OR, and XOR are supported). The more typical full 16-bit ALU implementation supports signed and unsigned arithmetic, a more complete set of logic operations (AND/OR/XOR/NOT/cNOT/BVS), full 16-bit left and right shifts, population count, as well as max/min functions. The resource cost is approximately double that of the minimum ALU. The 5 stage pipeline 32-bit version again doubles the logic, as might be expected, but the number of registers triples, as individual functions (specifically the adder/subtractor and shifters) are themselves pipelined to ensure that the ALU always exceeds 800MHz. This contrasts with the 16-bit ALU, where the pipelining is used to improve the placement of the entire ALU, rather than improving the performance of any individual function. There is also a 4 stage pipeline version of the 32-bit integer ALU, which is about the size of the 16-bit full function ALU. This returns a lower performance (typically 700 MHz), and is used in order to save logic for the QP version of the eGPU (which has a lower target speed of 600MHz). The individual resource counts in Table~\ref{tab:alu_fitting} may not accurately reflect the impact of each function to the overall ALU size, as synthesis may combine aspects of some functions together.

\subsection {Predicate Resources}

In Table~\ref{tab:dp_fitting} and ~\ref{tab:qp_fitting} the area impact of predicate support is clearly visible, increasing the soft logic resources by about 50\%. While each predicate stack (including its control) is very small, each thread has a unique stack. The base predicate area consists of only a thread comparator (which checks that the SP currently executing the thread that the predicate circuit is associated with), an instruction decode (IF/ELSE/ENDIF), and the single bit-wide predicate stack. This may only be 5 ALMs per thread, but if a typical eGPU contains 1$k$ threads, the predicate circuitry can quickly grow to be as large as the rest of the soft logic. Increasing the stack depth will have only a minimal impact on area, as each additional level consists of only a two input mux and a register.

\subsection {Instruction Fetch, Decode, and Control}

This section will always have a modest footprint, requiring 200 to 250 ALMs, and a handful of M20Ks to store the instruction words. The instruction decoder takes about 40 ALMs, and the thread generator around 25 ALMs. A single M20K can store 512 40-bit instruction words; the benchmarks we analyse later in this paper range from 30 instructions (32 element reduction) to 250 instructions (256 element bitonic sort), so a multi-tenancy of programs would only need several M20Ks. 

Increasing the IW to 43 or 46 bits (which is required to support a 32 and 64 registers per thread, respectively), adds only a single M20K per 2$k$ instructions, as the M20K containing the upper bits would be configured in $x$8 format. In any case, the number of M20Ks needed for program storage is small compared to the thread registers and shared memory. For example, a 1$k$ word program space would require three M20Ks, and a 4$k$ program space nine M20Ks. 

\subsection{Calculating Resources and Setting Performance}

Although the eGPU has a parameterized pipeline depth between the SPs and shared memory, it can achieve the target performance (771MHz and 600MHz respectively) using the minimum depth of 8 stages. The parameterized pipelining can be used for future applications with larger shared memories, or when the shared memories are placed elsewhere on the device, and not located near the SP array. We also report the slowest path outside the embedded (M20K and DSP) resources (see Table~\ref{tab:dp_fitting} and ~\ref{tab:qp_fitting}). If needed, there are also additional pipelining parameters inside the SP for the paths both to and from the FP and Integer ALUs. We will show in the next section how additional pipelining may not improve Fmax as the eGPU has been designed to fit into an Agilex sector in the base configuration. 

We can see that the SP overhead (mux and control) is $\approx$150 ALMs, the integer ALU ranges from $\approx$100 ALMs to $\approx$400 ALMs, and the predicates, if used, start from $\approx$ 150ALMs. A single SP will therefore be as small as 250 ALMs, and can be as large as 650 ALMs; this translates into a small eGPU core (16 SPs) requiring 4$k$ ALMs, and over 10$k$ ALMs for fully featured example. 

The number of M20Ks for the register memory for the DP eGPU can be calculated as $\text{threads} \times \text{registers}/256$; for the shared memory the number of blocks is $2 \times \text{size}(kB)$. The number of M20K blocks required for the QP eGPU are half of this, except that there is a minimum size ($\text{threads} \times \text{registers\_per\_thread}/16 >$ 2047) for the number of registers, in which case the QP eGPU will need the same number of register blocks as the DP version. 

\subsection{FPGA Sector Resources and Impact}

It is most beneficial to select eGPU parameters around the available FPGA resources and their on-chip organization. The Intel Agilex devices are arranged in sectors, the most common of which contains about 16400 ALMs, 240 M20K memories, and 160 DSP Blocks. Although we are not limited to a single sector (additional pipelining may be required to maintain performance across sector boundaries), this ratio of resources provides a good guide how to parameterize a eGPU instance. In particular, creating too large a register or memory space will be inefficient, as the ALMs between the M20K columns will likely be unreachable by other designs in the FPGA. Likewise, there is no point in specifying a small register space or shared memory, as the M20Ks between the logic structures may not be accessible by other functions. Further analysis is provided in the following section where we demonstrate that by selecting parameters in this way, the eGPU consistently achieves the reported performance levels by matching its architecture with the sector structure.

\begin{table*}
\small
  \begin{center}
    \caption{Fitting Results - DP Memory}
    \label{tab:dp_fitting}
    \begin{tabular}{|c|c|c|c|c|c|c|c|c|c|c|c|c|}
    \hline
  \multirow{2}{*}{\textbf{Scale}} & \textbf{ALU} & \textbf{Shift} & 
  \multirow{2}{*}{\textbf{Threads}} & \textbf{Reg.} & \textbf{Shared} & \textbf{Predicate} & \multirow{2}{*}{\textbf{ALM}} & \multirow{2}{*}{\textbf{Registers}} & \multirow{2}{*}{\textbf{DSP}} & \multirow{2}{*}{\textbf{M20K}} & \textbf{Freq} & \textbf{SP}\\   
  
   & \textbf{Precision} & \textbf{Precision} & \textbf{} & \textbf{Thread} & \textbf{Memory} & \textbf{Levels} & & & & & \textbf{(MHz)} & \textbf{(ALM/Reg.)}  \\
      \hline\hline
    \ {Small} & {16} & {1} & {512} & {16} & {8KB} & {0} & {4243} & {13635} & {24} & {50} & {1018/771} & {224/707} \\
    \hline
    \ {Small} & {16} & {16} & {512} & {16} & {32KB} & {5} & {7518} & {18992} & {24} & {98} & {898/771} & {413/979}  \\
    \hline
    \ {Medium} & {16} & {16} & {512} & {32} & {32KB} & {5} & {7579} & {19155} & {24} & {131} & {883/771} & {426/1043}  \\
    \hline
    \ {Medium} & {32} & {16} & {512} & {32} & {32KB} & {5} & {9754} & {25425} & {24} & {131} & {902/771} & {461/1277}  \\
    \hline
    \ {Large} & {32} & {16} & {512} & {64} & {32KB} & {8} & {10127} & {26040} & {32} & {195} & {860/771} & {575/1505}  \\
    \hline
    \ {Large} & {32} & {32} & {512} & {64} & {64KB} & {16} & {10697} & {26618} & {32} & {259} & {841/771} & {600/1476}  \\
    \hline
    \end{tabular}
  \end{center}
\end{table*}

\begin{table*}
\small
  \begin{center}
    \caption{Fitting Results - QP Memory}
    \label{tab:qp_fitting}
    \begin{tabular}{|c|c|c|c|c|c|c|c|c|c|c|c|c|c|}
    \hline
  \multirow{2}{*}{\textbf{Scale}} & \textbf{ALU} & \textbf{Shift} & 
  \multirow{2}{*}{\textbf{Threads}} & \textbf{Regs./} & \textbf{Shared} & \textbf{Predicate} & \multirow{2}{*}{\textbf{ALM}} & \multirow{2}{*}{\textbf{Registers}} & \multirow{2}{*}{\textbf{DSP}} & \multirow{2}{*}{\textbf{M20K}} & \textbf{Freq} & \textbf{SP} \\   
   & \textbf{Precision} & \textbf{Precision} & \textbf{} & \textbf{Thread} & \textbf{Memory} & \textbf{Levels} & & & & & \textbf{(MHz)} & \textbf{(ALM/Reg.)}   \\
      \hline \hline
    \ {Small} & {32} & {1} & {512} & {64} & {32KB} & {0} & {5468} & {14487} & {24} & {98} & {840/600} & {287/830} \\
    \hline
    \ {Medium} & {32} & {32} & {1024} & {32} & {64KB} & {0} & {7057} & {16722} & {32} & {131} & {763/600} & {396/1016} \\
    \hline
        \ {Large} & {32} & {32} & {1024} & {32} & {64KB} & {16} & {11314} & {25050} & {32} & {131} & {763/600} & {685/1601} \\
    \hline
        \ {Large} & {32} & {32} & {1024} & {32} & {128KB} & {10} & 
        {10174} & {23094} & {32} & {195} & {714/600} & {556/1391} \\

        \hline
        \end{tabular}
  \end{center}
\end{table*}

\begin{table}
\footnotesize
  \begin{center}
    \caption{Fitting Results - Integer ALU}
    \label{tab:alu_fitting}
    \begin{tabular}{|c|c|c|c|c|c|c|c|c|}
    \hline
  \multirow{2}{*}{\textbf{Prec.}} & \multirow{2}{*}{\textbf{Type}} & \multirow{2}{*}{\textbf{ALM}} & \multirow{2}{*}{\textbf{Registers}} &\textbf{Add/} & \multirow{2}{*}{\textbf{Logic}} & \multirow{2}{*}{\textbf{SHL}} & \multirow{2}{*}{\textbf{SHR}} & \multirow{2}{*}{\textbf{Pop}} \\
  \textbf{} & \textbf{} & &  & \textbf{Sub} & \textbf{} & \textbf{} & \textbf{} & \textbf{} \\
  \hline \hline
  \ {16} & {Min} & {90} & {136} & {3} & {9} & {-} & {-} & {-}\\
  \hline
  \ {16} & {Small} & {134} & {207} & {9} & {10} & {20} & {23} & {-}\\
  \hline
  \ {16} & {Full} & {199} & {269} & {9} & {18} & {20} & {23} & {11}\\
  \hline
  \ {32} & {Min} & {208} & {406} & {5} & {27} & {28} & {28} & {-}\\
  \hline
  \ {32} & {Full} & {394} & {704} & {27} & {36} & {50} & {53} & {27}\\
  \hline
  \end{tabular}
  \end{center}
\end{table}

%\newpage\clearpage
%%%%%%%%%%%%%%%%%%%%%%%%%%%%%%%%%%%%%%%%%%%%%%%%%%%%%%%%%%%%%%%%%%%%%%%%%%%
\section{Repeatable High Performance}
%%%%%%%%%%%%%%%%%%%%%%%%%%%%%%%%%%%%%%%%%%%%%%%%%%%%%%%%%%%%%%%%%%%%%%%%%%%

This section provides the required information to make our design process repeatable for those wishing to achieve high performance in their own designs. It is therefore necessarily `close to metal' in abstraction. Although the details are specific to Intel FPGAs, we believe the same approaches are valid for all other FPGAs as well.

The eGPU is designed to give consistent performance, which will always be limited by the slowest embedded (DSP or M20K memory) resource in that configuration. The clock network in Agilex is specified at 1GHz, which is the absolute limit of performance for any design. The M20K memories in DP mode also achieve 1GHz, but only 600 MHz in QP mode. The DSP Blocks can run at 771 MHz when implementing a FP32 multiply-add datapath with a 4 stage pipeline~\cite{DSPUG}.
We are therefore limited to a maximum speed of 771MHz, unless we use QP memory, in which case the maximum frequency drops to 600MHz. The lower performance of the QP memory, however, will allow us to support a higher density storage, and the doubled write bandwidth may offer an overall higher throughput for some applications. We will examine some of these trade-offs in the benchmarking section.

Using the sector architecture effectively enables the eGPU performance and efficiency. Sector resources are arranged in columns, each approximately 41 rows high (several columns are shorter because of non-user accessible device features). Achieving a 1GHz speed for soft logic does not require logic to be immediately adjacent to each other, as there are different horizontal and vertical wire lengths - too much pipelining can negatively impact performance as much as too little. More important is using the minimal number of wire resources per connection. In the Agilex devices, there is a constant 4 columns of logic between each column of either DSP or M20K. In a sector we will have 40 columns of logic, 4 columns of DSP, and 6 columns of M20K. There is little point in saving logic or memory if it is not accessible by other portions of design.

As we have shown in the previous section, the results are deterministic and repeatable, in both area and performance. Ideally, the resource use would be balanced to realize the maximum efficiency from the device.

To map eGPU to the device, we first sketched out a LUT level construct of an SP, and adjusted it so that the number of logic levels would align with the sector column ratios described above. Paths directly between M20K memories (which implement the thread registers in each SP) and the DSP Blocks had to fit into a 4 column group of LABs, and longer pure logic paths ({\em{e.g.}} the integer ALUs) were organized so that the total area did not spill over into a M20K or DSP column that might be used by another SP.  

We can see from the results (Table~\ref{tab:alu_fitting}) that a 16-bit Integer ALU is in the range of 100-200 ALMs and the 32-bit version requires 200-400 ALMs. If predicates are used, they will cost an additional 125-250 ALMs per SP, depending on the defined thread space. The remaining 150 ALMs per SP are used for the data muxing and alignment shown in Figure~\ref{fig:SM_SP_Top}. We were able to implement a small eGPU (the first example in Table~\ref{tab:dp_fitting})  that was able to close timing over 771 MHz with no dedicated soft logic registers ({\em{i.e.}~}registers that were not directly connected with a logic function, such as a mux), but for the generic parameterized case, we added a single additional pipeline stage register between the thread registers and the functional units, and also one level in the write-back path between the functional units and the thread registers. For all of the examples in Table~\ref{tab:dp_fitting} there are also single pipeline stages to and from the shared memory. We parameterized the pipeline depth for all of these stages, along with the appropriate balancing delays for the data and control paths into the write paths of the thread registers, but found that these were not needed to be increased beyond one pipeline stage for any of the reported examples.

\begin{figure}
    \centering
    \includegraphics[scale=0.6]{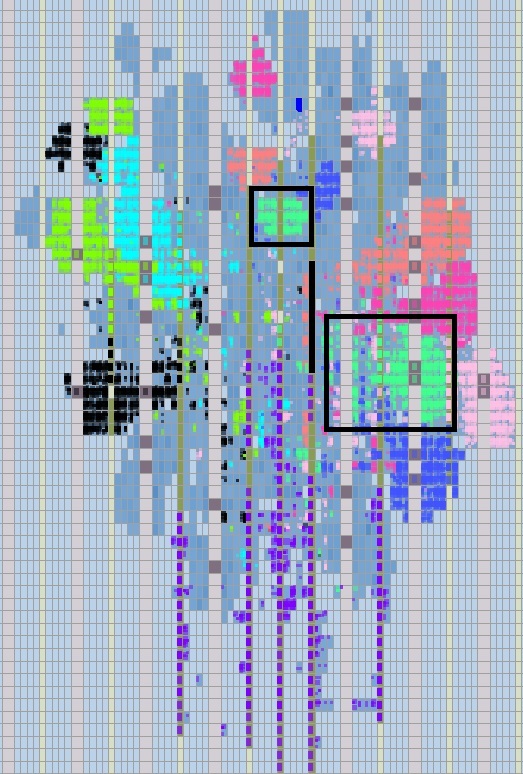}
    \caption{eGPU Placement}
    \label{fig:eGPU_place}
\end{figure}

Figure~\ref{fig:eGPU_place} shows the unconstrained placement of the largest instance of Table~\ref{tab:dp_fitting}. The shared memory and 8 out of the 16 SPs have been color coded for identification. The shared memory creates a spine in the middle of the core, with 8 SPs placed on either side of it. For purposes of illustration we have colored a subset of SPs: three random SPs and the left of the spine, and five contiguous ones on the right. Three things are evident with all SPs: (a) the majority of the logic is in one contiguous block, (b) there is a separate contiguous structure (the predicate block) placed some distance away, and (c), the SP straddles a columns of DSP Blocks. All of the instances of Table~\ref{tab:dp_fitting} and~\ref{tab:qp_fitting} display this pattern, including the shared memory spine. 

\begin{figure}
    \centering
    \includegraphics[scale=0.72]{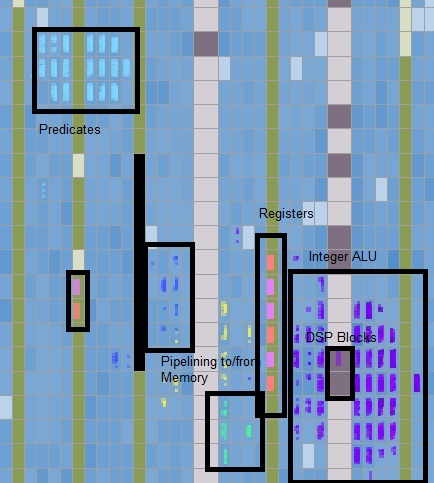}
    \caption{Single SP Placement}
    \label{fig:SP_place}
\end{figure}

Figure~\ref{fig:SP_place} shows one of the SPs in greater detail (this SP is the one marked by the black boxes in Figure~\ref{fig:eGPU_place}). The largest component is the integer ALU. The operators (adder/subtractor, shifters, arithmetic logic, {\em etc.}) are in the 4 columns to the right of the two DSP blocks (the DSP Block for the floating point operators is adjacent to the integer multiplier). To the left of the DSP Blocks is largely pipelining logic - of the 5 pipeline stages in the ALU, only one is used for pipelining the operators - the rest is used to break up the paths between the thread register memories and the ALUs. We examined all of the SP placements, and the placement of the M20Ks for the register memories (8 M20Ks for this instance) was in one of three layouts: (a) a contiguous single column (b) most of the registers in one column, with a smaller number in the next column further away from the integer ALU, and (c) equally split between two columns on either side of the integer ALU. In all of these cases, the pipeline wrapper around the ALU was usually grouped together, and essentially separate from the actual operators. Rather than having to be in a specific location relative to the M20Ks and operators, the ability to split up a bus so that it can be mapped to the same number of wire hops is what was important. Fewer pipeline stages would have introduced a two stage routing path, which would have likely become the critical path in the eGPU. (In the QP memory version, we can remove one of the pipeline stages as the M20K becomes the slowest component at 600 MHz, and we can see that the removal of some of the pipeline path reduce the non-memory path performance to just over 700 MHz). On the other hand, more than a 5 stage integer ALU could potentially decrease performance as it could spread out the placement of the SP.   

The predicate circuitry is placed in another contiguous block, but well away from the SP core it is associated with. From Figure~\ref{fig:eGPU_place} we can see that the majority of the other predicates have a similar relationship with their respective SP. All of these have been automatically placed by Quartus. This is possible because the interface to and from the predicate block is very narrow, with only a single bit ({\texttt{thread\_active}} signal) returned. The signals to the block are relatively few: a thread index (typically 5 to 8 bits wide), a 3-bit decoded instruction signal (IF/ELSE/ENDIF), and a single bit valid condition code. Although there are many possible conditions from many different instructions, these can be decoded into a single valid condition bit in the main SP body. These narrow busses give us flexibility to wrap multiple pipes around the relatively simple (consisting largely of a chained registers organized in individual stacks) predicate blocks, which makes it possible for the tool to place them almost completely independently of the main datapaths.

To create repeatable high performance designs, we need to understand both the structure, and the position of embedded features to each other. Here we are using integer ALUs which range in size by four times, our logic and memory density is very high, but our performance always exceeds that of the slowest embedded feature. It is possible to build a completely different type of CPU (or indeed any other type of core) and achieve this type of performance via a push button flow, but the architecture of the FPGA needs to be considered at every stage of the IP architecture phase. 

%%%%%%%%%%%%%%%%%%%%%%%%%%%%%%%%%%%%%%%%%%%%%%%%%%%%%%%%%%%%%%%%%%%%%%%%%%%
\section{Benchmarks}
%%%%%%%%%%%%%%%%%%%%%%%%%%%%%%%%%%%%%%%%%%%%%%%%%%%%%%%%%%%%%%%%%%%%%%%%%%%

\begin{table*}
\footnotesize
  \begin{center}
    \caption{Vector and Matrix Benchmarks}
    \label{tab:matrix_bench}
    \begin{tabular}{|c|c||c|c|c|c||c|c|c||c|c|c|c|c|}
    \hline
                     &                 & \multicolumn{4}{c||}{\bf Vector Reduction}  & \multicolumn{3}{c||}{\bf Matrix Transpose} & \multicolumn{5}{c|}{\bf Matrix x Matrix} \\
    \cline{3-14}
  \textbf{Dimension} & \textbf{Metric} & \multirow{2}{*}{\textbf{Nios}} & \textbf{eGPU} & \textbf{eGPU} & \textbf{eGPU} & \multirow{2}{*}{\textbf{Nios}} & \textbf{eGPU} & \textbf{eGPU} & \multirow{2}{*}{\textbf{Nios}} & \multirow{2}{*}{\textbf{FlexGrip}} & \textbf{eGPU} & \textbf{eGPU} & \textbf{eGPU}\\
  \ & & & \textbf{DP} & \textbf{QP} & \textbf{Dot} & & \textbf{DP} & \textbf{QP} & & & \textbf{DP} & \textbf{QP} & \textbf{Dot}  \\
      %\multicolumn{2}{*}{Type = INT, UNIT, FP32}\\
      \hline\hline
      \multirow{5}{*}{32} & {Cycles} & {459} & {168} & {160} & {62} & {21809} & {1720} & {1208} & {1.45M} & {2.14M} & {111546} & {103354} & {19800} \\ %1454394
    \ & {Time(us)} & {1.32} & {0.22} & {0.27} & {0.08} & {62.85} & {2.23} & {2.01} & {4179} & {21400} & {144.7} & {172.3} & {25.7} \\
    \ & {Ratio(cycles)} & {2.73} & {1.0} & {0.95} & {0.37} & {12.68} & {1.0} & {0.7} & {13.03} & {19.2} & {1.0} & {0.93} & {0.18} \\
    \ & {Ratio(time)} & {6.01} & {1.0} & {1.23} & {0.37} & {28.18} & {1.0} & {0.9} & {28.97} & {147.9} &{1.0} & {1.19} & {0.18} \\
    \ & {Normalized} & {1.14} & {1.0} & {1.4} & {0.45} & {5.33} & {1.0} & {1.02} & {5.48} & {-} & {1.0} & {1.35} & {0.21} \\
    \hline
    \multirow{5}{*}{64} & {Cycles} & {1803} & {202} & {194} & {94} & {86609} & {5529} & {3481} & {11.6M} & {16.6M} & {451066} & {418671} & {84425} \\ %11584090
    \ & {Time(us)} & {5.20} & {0.26} & {0.32} & {0.12} & {249.6} & {7.17} & {5.80} & {33383} & {166000} & {585.0} & {697.8} & {109.5} \\
    \ & {Ratio(cycles)} & {8.93} & {1.0} & {0.96} & {0.47} & {15.66} & {1.0} & {0.63} & {25.7} & {36.8} & {1.0} & {0.93} & {0.19} \\
    \ & {Ratio(time)} & {19.98} & {1.0} & {1.23} & {0.47} & {34.81} & {1.0} & {0.81} & {57.1} & {284} & {1.0} & {1.19} & {0.19} \\
    \ & {Normalized} & {3.78} & {1.0} & {1.4} & {0.60} & {6.59} & {1.0} & {0.92} & {10.80} & {-} & {1.0} & {1.35} & {0.23} \\
    \hline
    \multirow{5}{*}{128} & {Cycles} & {3595} & {216} & {208} & {101} & {345233} & {20481} & {12649} & {92.5M} & {441.2M} & {2342356} & {2212136} & {886452} \\ %92472474
    \ & {Time(us)} & {10.36} & {0.28} & {0.35} & {0.13} & {994.91} & {26.56} & {21.08} & {266491} & {4412.1} & {3038.1} & {3686.9} & {1149.7} \\
    \ & {Ratio(cycles)} & {16.64} & {1.0} & {0.96} & {0.47} & {16.86} & {1.0} & {0.62} & {39.47} & {188.3} & {1.0} & {0.94} & {0.38} \\
    \ & {Ratio(time)} & {37.00} & {1.0} & {1.23} & {0.47} & {37.45} & {1.0} & {0.79} & {87.71} & {1452} & {1.0} & {1.21} & {0.38} \\
    \ & {Normalized} & {7.00} & {1.0} & {1.4} & {0.60} & {7.09} & {1.0} & {0.90} & {1659} & {-} & {1.0} & {1.37} & {0.46} \\
    \hline
    \end{tabular}
  \end{center}
\end{table*}

\begin{figure*}
    \centering
    \includegraphics[scale=0.28]{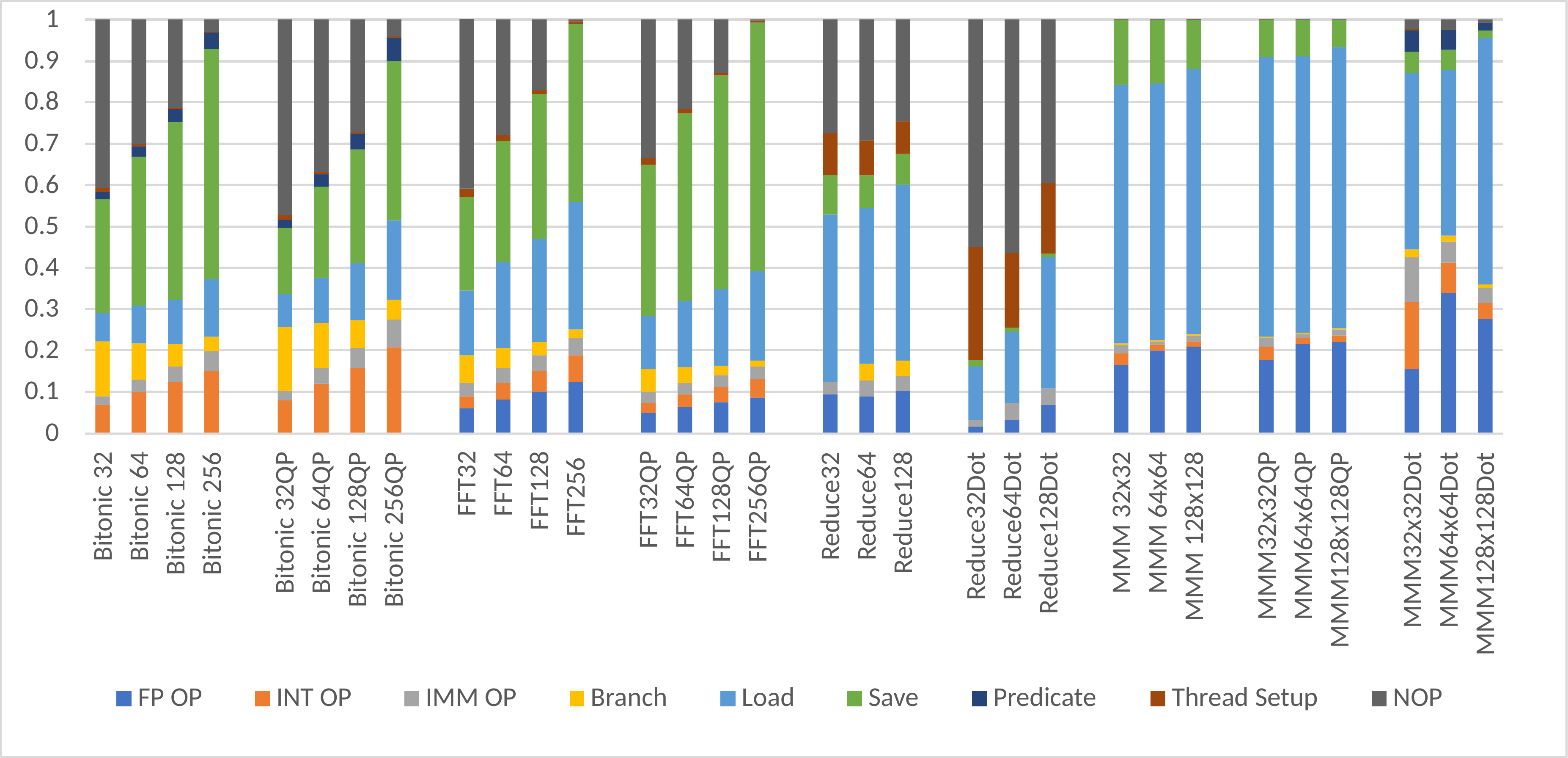}
    \caption{Benchmark Profiling (Y-Axis shows proportion of instructions executed by type).}
    \label{fig:MMM_profile}
\end{figure*}

We ran a number of workloads of different types to evaluate absolute and relative performance of the eGPU for varying data sizes that we might expect for embedded applications. We also profiled all the workloads to examine the efficiency of the eGPU. For continuity we selected many of the same benchmarks as used by Flexgrip~\cite{FlexGrip_Thesis}. We chose vector reduction, matrix transpose, and matrix-matrix multiply (MMM), as these would be common building blocks for many GPGPU applications. Bitonic sort~\cite{bitonic} is a sorting algorithm suited for parallel processing. Instead of the simpler autocorrelation, we used the FFT, as we felt this would be more representative of the workloads expected for the eGPU. All benchmarks were written in assembly code (we have not written our compiler yet). 

We report the comparison to FlexGrip only for the MMM, as the larger dataset size would be less affected by any overheads for setup and data transfer. We see that there is a significant performance advantage in favor of eGPU in cycle time alone. We ran all benchmarks, except the FFT, for which there are no reported FlexGrip results. FlexGrip underperforms eGPU by a factor of $\approx$31x, averaged over all benchmarks.  We did not compare against DO-GPU (which is the latest iteration of FGPU), as DO-GPU normalized size is 50x-100x greater than eGPU.

Our reported measurements are all based on core performance: we start the clock once the data has been loaded into the shared memory, and stop the clock once the final result has been written back to the shared memory. The most likely use of the eGPU is to apply multiple algorithms to the same data - $i.e.$ there is no loading and unloading of data between different algorithms. For completeness, we also ran all of our benchmarks taking into account the time to load and unload the data over the 32-bit wide data bus. The performance impact was only 4.7\%, averaged over all benchmarks. 

Clock frequency was 771 MHz for eGPU (including where the Dot Product operator is used), and 600 MHz for eGPU-QP variant. We compare both cycle counts and elapsed time for the eGPUs with the two shared memory architectures, and also the impact of the optional Dot Product core for reduction and MMM benchmarks. We then normalize the performance (time), by the resource cost, which we calculated on the basis of ALMs and DSP Blocks. We estimate that the effective cost of a DSP block is 100 ALMs, which we calculate as follows: we start with the ALM count of the pure soft logic implementation of a FP32 multiply and adder (approximately 650 ALMs~\cite{FPUG}), and add 50\% area to this number for DSP Block overhead (a DSP Block contains considerable additional features). We then divide by 10 for an approximate soft logic to hard logic scaling factor (earlier work~\cite{KuonRose} suggested a higher ratio in the general case, but recent work~\cite{arithfpgamult} described more efficient ways of mapping arithmetic, especially multipliers, to FPGAs). We report normalized cost (considering both elapsed time and resources, with eGPU-DP as the baseline).

As a comparison, we ran all of the benchmarks on Nios IIe~\cite{Nios}, which is a mature RISC processor for Intel FPGAs. The configuration we used consumed 1100 ALMs (plus 3 DSP Blocks, giving a normalized cost of 1400), and closed timing at 347 MHz. We did not profile the Nios code, but analyzed the efficiency of operation (CPI). Most of the benchmarks retired an instruction every 1.7 clock cycles, except for the matrix-matrix multiplies and FFT, which required about 3 clocks, because of the way that 32$\times$32 multipliers were implemented. (For simplicity, we replaced the FP32 arithmetic with INT32 for the Nios examples). 

For the vector and matrix benchmarks, we chose an eGPU configuration with 32 registers per thread, with a 32 bit ALU, and a 128KB shared memory. This configuration has an equivalent cost (see Table~\ref{tab:dp_fitting} and~\ref{tab:qp_fitting}) of 7400, 8400, and 9000 ALMs for the eGPU-DP, eGPU-QP, and eGPU-Dot variants respectively. 
Depending on the configuration, eGPU is $5\times$ to $6\times$ larger than Nios (but also more than twice the operating frequency). We would therefore expect (or at least hope) that eGPU would give an OOM performance increase over Nios.

We can deduce the mechanism of the matrix transpose benchmarks from Table~\ref{tab:matrix_bench} directly. For a given $n \times n$ matrix, we know that the eGPU will need {$n^2$} cycles to write the transposed elements to shared memory and 1/4th of those cycles to initially read them into the SP threads. We can see that the number of cycles clocked is marginally larger than this; these are largely used for the integer instructions needed to generate the transposed write addresses. We expect that the eGPU-QP will require about 40\% fewer cycles, being able to write two transposed elements per clock, which indeed is the case.

 The vector reduction needs inter-SP communication, which go through the shared memory, which is the performance bottleneck in the eGPU. Table~\ref{tab:matrix_bench} shows the impact of memory accesses on reduction performance. The actual floating point operations are a relatively small ($\approx$10\%) component of the reduction, with the majority of the cycles used by the memory operations. If we are using the dot product operator, there are even fewer FP operations required, and most of the time is spent waiting (NOPs) for the dot product to write back to the SP. All final vector reductions end up in the first SP, and we can use the multi-threaded CPU or MCU eGPU dynamic scaling personalities to write these values to the shared memory. 

The MMMs are much more complex. Although the algorithm itself is very simple, consisting only of a three level loop, the standard GPU implementation requires a vector reduction.
While the cycle count increases as expected ($\sim4\times$) from 32$\times$32 to 64$\times$64, there is an unexpected jump from 64$\times$64 to 128$\times$128, which is particularly evident in the eGPU-Dot case. Analysis of the code shows that while we are able to store the entire matrix (or at least a majority of the matrix) in the SP registers (there are 16384 total registers across the 16 SPs in the configuration we have chosen here) for the 32$\times$32 and 64$\times$64 cases, we need to keep reloading portions of the matrix in the 128$\times$128 case, which can also be seen in the profile stack in Figure~\ref{fig:MMM_profile}. Of course, there is always the option of increasing the maximum thread space or registers per thread (through parameterization) if the expected workloads were larger matrices. Compared to the vector reduction (where we profile a single vector), the thread initialization and integer operations are amortized away as we operate on many vectors. The NOPs also disappear as the the thread depth increases here.

The bitonic sort benchmark requires a wider mix of instructions. Predicates are required, which increases the effective cost of the eGPU core by about 50\%. The smaller sorts require many NOPs, which progressively reduce as the number of wavefronts increase for the larger datasets. The nature of the bitonic sort tends to use many subroutine calls, which we can see here in the relatively large number of branch operations. Again, the memory operations take the majority of all cycles, as each pass of the sort requires a redistribution of the data among the SPs. While the eGPU-QP version requires fewer clock cycles because of the increased write bandwidth, the normalized cost of the QP version is higher, largely because of the lower clock frequency.

A similar pattern of instruction distribution is seen in the FFT. Increasing wavefront depth for larger datasets reduces NOPs significantly. The number of FP instructions (which are doing the actual FFT calculations) is relatively small, at about 10\%. The largest proportion of operations are once again the memory accesses, especially in the write to shared memory; using the QP version of the eGPU results in a 20\% to 30\% decrease in total cycles. The normalized cost of the two eGPU versions, however, is approximately the same, with the high clock frequency of the base version offsetting the higher memory bandwidth of the QP version. These results also point to a better optimization for the FFT: by using a higher radix FFT, there will be correspondingly fewer passes through the shared memory. (We have a extensive flexibility in specifying the register and thread parameters, we can easily support much higher radices, which will require much larger register spaces). 

Comparing against Nios, we can see that the eGPU performs very well. We see at least an OOM performance difference based on time, and in almost all cases on a cycle basis as well. This tells us that eGPU is a more efficient architecture than a RISC processor, and is a viable candidate for a soft accelerator core.

\begin{table}
\footnotesize
  \begin{center}
    \caption{Bitonic Sort and FFT Benchmarks}
    \label{tab:bitonic_bench}
    \begin{tabular}{|@{~}c@{ }|c||c|c|c||c|c|c|}
    \hline
               &                 &  \multicolumn{3}{c||}{\bf Bitonic Sort}  &  \multicolumn{3}{c|}{\bf FFT}  \\
  \cline{3-8}             
  \textbf{Dim} & \textbf{Metric} & \multirow{2}{*}{\textbf{Nios}} & \textbf{eGPU} & \textbf{eGPU} & \multirow{2}{*}{\textbf{Nios}} & \textbf{eGPU} & \textbf{eGPU}\\
               &                 &               &      \bf DP         & \bf QP        &               &     \bf DP          & \bf QP        \\
      %\multicolumn{3}{*}{Type = INT, UNIT, FP32}\\
      \hline\hline
      \multirow{5}{*}{32} & {Cycles} & {8457} & {1742} & {1543} & {9165} & {876} & {714} \\
    \ & {Time(us)} & {24.37} & {2.25} & {2.51} & {26.41} & {1.14} & {1.19}  \\
    \ & {Ratio(cycles)} & {4.89} & {1.0} & {0.86} & {10.46} & {1.0} & {0.82} \\
    \ & {Ratio(time)} & {10.8} & {1.0} & {1.1} & {23.16} & {1.0} & {1.04} \\
    \ & {Normalized} & {1.24} & {1.0} & {1.24} & {4.38} & {1.0} & {1.18}\\
    \hline
    \multirow{5}{*}{64} & {Cycles} & {20687} & {3728} & {3054} & {20848} & {1695} & {1312} \\
    \ & {Time(us)} & {59.6} & {4.83} & {5.09} & {60.08} &  {2.20} & {2.19}  \\
    \ & {Ratio(cycles)} & {5.54} & {1.0} & {0.82} & {12.30} &  {1.0} & {0.82} \\
    \ & {Ratio(time)} & {12.3} & {1.0} & {1.05}  & {27.31} &  {1.0} & {1.01} \\
    \ & {Normalized} & {1.42} & {1.0} & {1.18}  & {5.17} &  {1.0} & {1.13}\\
    \hline
    \multirow{5}{*}{128} & {Cycles} & {49741} & {8326} & {6536} & {46667} & {3463} & {2558} \\
    \ & {Time(us)} & {143.3} & {10.8} & {10.9} & {134.49} &  {4.29} & {4.26}  \\
    \ & {Ratio(cycles)} & {5.97} & {1.0} & {0.79} & {13.48} &  {1.0} & {0.74} \\
    \ & {Ratio(time)} & {13.2} & {1.0} & {1.01}  & {31.35} &  {1.0} & {0.95} \\
    \ & {Normalized} & {1.48} & {1.0} & {1.13} & {5.93} &  {1.0} & {1.08}\\
    \hline
    \multirow{5}{*}{256} & {Cycles} & {149271} & {16578} & {11974} & {103636} & {6813} & {4736} \\
    \ & {Time(us)} & {430.2} & {21.5} & {19.9} & {298.66} &  {8.84} & {7.89}  \\
    \ & {Ratio(cycles)} & {9.0} & {1.0} & {0.72} & {15.21} &  {1.0} & {0.70} \\
    \ & {Ratio(time)} & {20.0} & {1.0} & {0.93}  & {33.79} &  {1.0} & {0.89} \\
    \ & {Normalized} & {2.24} & {1.0} & {1.05} & {6.39} &  {1.0} & {1.01}\\
    \hline
    \end{tabular}
  \end{center}
\end{table}

%%%%%%%%%%%%%%%%%%%%%%%%%%%%%%%%%%%%%%%%%%%%%%%%%%%%%%%%%%%%%%%%%%%%%%%%%%%
\section{Conclusions}
%%%%%%%%%%%%%%%%%%%%%%%%%%%%%%%%%%%%%%%%%%%%%%%%%%%%%%%%%%%%%%%%%%%%%%%%%%%
We have demonstrated a GPGPU that consistently beats 770 MHz for a wide range of parameters, and described the design approach required to reach such frequencies. We are able to swap in and out features as well as change the precision of the integer ALU to optimize for area and resource balancing in the FPGA. 

For the eGPU to be useful in an actual system design, it must offer an improvement over known methods. We compare the eGPU to a mature  commercial soft CPU (Nios) over a number of benchmarks. The eGPU is much better on a cycle by cycle or elasped time basis in all cases we tried (typically by one to two OOM), and is still better on an area normalized basis. When we add the dot product core - which can be used directly by the eGPU in a regular GPGPU context - the advantage can increase again by several times. A soft GPU therefore can offer a valid implementation option for many types of algorithms. This does not mean that a GPGPU will replace the RISC, anymore than a discrete GPGPU will replace a discrete RISC, only that we have shown that the soft GPGPU can now be considered for commercial designs, rather than just being of academic interest. The eGPU only uses 1\%-2\% of a current mid-range device, making it a cost effective option to implement complex algorithms in a larger FPGA system design, even if multiple cores are required.

\newpage\clearpage
\balance
\bibliographystyle{ACM-Reference-Format}
\bibliography{references}

\end{document}